\documentclass[]{spie}

\usepackage[]{graphicx}
\usepackage{subfigure}
\usepackage{color}
\usepackage{amsmath}
\usepackage{hyperref}

\title{Gemini Planet Imager Observational Calibration XIII: Wavelength Calibration Improvements, Stability, and Nonlinearity} 

\author{Schuyler G. Wolff\supit{a},  Kimberly Ward-Duong\supit{b}, Joe Zalesky\supit{c}, Alex Z. Greenbaum\supit{a}, Marshall D. Perrin\supit{d}, and James Graham\supit{c}, with the GPI team.
\skiplinehalf
\supit{a}Johns Hopkins University, 3400 North Charles St., Baltimore, MD, USA; \\
\supit{b}School of Earth and Space Exploration, Arizona State University, PO Box 871404, Tempe, AZ 85287, USA; \\
\supit{c}Astronomy Department, University of California, Berkeley; Berkeley, CA 94720, USA; \\
\supit{d}Space Telescope Science Institute, 3700 San Martin Drive, Baltimore, MD, USA;  \\
}

\authorinfo{Further author information: (Send correspondence to S.G.W.)\\  S.G.W.: E-mail: swolff@stsci.edu, Telephone: 1 410 338 2428 }

\begin{document} 
  \maketitle 

\begin{abstract}
We present improvements to the wavelength calibration for the lenslet-based Integral Field Spectrograph (IFS), that serves as the science instrument for the Gemini Planet Imager (GPI). The GPI IFS features a $2.7'' \times 2.7''$ field of view and a 190 x 190 lenslet array (14.1 mas/lenslet) with spectral resolving power ranging from R $\sim$ 35 to 78. A unique wavelength solution is determined for each lenslet characterized by a two-dimensional position, an n-dimensional polynomial describing the spectral dispersion, and the rotation of the spectrum with respect to the detector axis. We investigate the non-linearity of the spectral dispersion across all \textit{Y, J, H,} and \textit{K} bands through both  on-sky arc lamp images and simulated IFS images using a model of the optical path. Additionally, the 10-hole non-redundant masking mode on GPI provides an alternative measure of wavelength dispersion within a datacube by cross-correlating reference PSFs with science images. This approach can be used to confirm deviations from linear dispersion in the reduced datacubes. We find that the inclusion of a quadratic term provides a factor of 10 improvement in wavelength solution accuracy over the linear solution and is necessary to achieve uncertainties of a few hundredths of a pixel in \textit{J} band to a few thousands of a pixel in the \textit{K} bands. This corresponds to a wavelength uncertainty of $\sim$ 0.2 nm across all filters.
\end{abstract}



\section{Gemini Planet Imager Wavelength Calibration}
\label{sec:intro}  

The Gemini Planet Imager (GPI) is a high contrast instrument located at Gemini South designed to directly detect and characterize exoplanets. The main science instrument on GPI is an Integral Field Spectrograph (IFS) capable of obtaining low resolution spectroscopy with R $\sim$ 35 - 78 across all the \textit{Y, J, H,} and \textit{K} bands. The GPI IFS uses a lenslet based design to reproduce 190 x 190 microspectra in the narrow 2.7'' x 2.7'' field of view of the Hawaii 2RG detector.\cite{2014AAS22320204C} Each bandpass is split into 37 wavelength channels, with \textit{K} band split into two separate bands to limit the overlapping of the microspectra for adjacent lenslets. The field of view and spectral resolution were carefully chosen to produce Nyquist sampling on the detector at the shortest wavelengths, while providing sufficient spectral resolution at the longer wavelengths to resolve important features in planetary atmospheric models.\cite{2011PASP123692M}  

A wavelength solution is determined for each lenslet individually, and is characterized by a 2D position, the dispersion, and the tilt (angle from vertical in radians) of the spectrum. Each set of parameters is different for the $\sim$ 36000 lenslets and vary widely across the field of view. A cutout of a single lenslet is modeled by placing 2D Gaussian PSFs at the predicted spectral peak locations for the GCAL (Gemini's calibration lamp unit)\cite{1997SPIE.2871.1171R}  Xe or Ar arc lamp sampled at the resolution of GPI. The data and model spectra are compared using the constrained Levenberg-Marquardt least-squares minimization algorithm to minimize the error weighted squared residuals of the two images over the free parameters above.\cite{Levenberg}\cite{marquardt} This allows measurement of spectral positions to better than 1/10th of a pixel. For a full description of this algorithm, see Wolff et al. (2014).\cite{wavecals}

In this paper, we present work completed on the wavelength calibration of the Gemini Planet Imager since 2014, and includes many lessons learned from $\sim$ 2 years on sky at Gemini South. In Section 1.1, we describe the recommended practices for the wavelength calibration of any spectral mode GPI science data for the general GPI user. In Section 1.2, we provide techniques for checking the quality of your data and troubleshooting any data quality issues. In Section 2, we discuss the stability of the wavelength calibration over time. In Section 3, we examine the non-linearity of the wavelength solution. Finally, we discuss ongoing work on the simultaneous wavelength solution across all filters. 

\subsection{Recommended Practices}

For the ease of use of the general observer, the GPI team has provided a publically available Data Reduction Pipeline (DRP) for the calibration and analysis of both spectral and polarimetry modes of GPI.\cite{2014AAS22334814P,perrinpipe} `Recipes' are furnished for different types of data and are made up of `Primitives' which handle the individual tasks (Ex. dark subtraction, bad pixel correction, wavelength calibration, generating the spectral datacubes, PCA post processing techniques etc.). There are several recipes devoted to wavelength calibration. These are dicussed in detail below. See Perrin et al. (these proceedings) for a full description of the data infrastructure developed for the GPI Exoplanet Survey Campaign, including the GPI DRP.  Below we discuss the steps for properly calibrating the spectra of a science target; both the data to be obtained, and the data reduction processes using the GPI DRP. 

\begin{itemize}

\item As part of Gemini South's regular calibration program, a set of high signal-to-noise arc lamp images are obtained each time GPI is remounted on the bottom port of the Gemini South telescope using the Gemini Facility Calibration Unit (GCAL). The Argon arc lamp is used as it requires a shorter exposure time than the Xenon lamp to acheive the same SNR. 

\item The user should reduce these data using the \textbf{Wavelength Solution 2D} Recipe in the GPI DRP. With the new release of the GPI DRP in July 2016, the user now has the option to choose between a Gaussian PSF and a Microlens PSF\cite{microlensspie} to simulate the arc lamp spectra. The Microlens PSF generally does a better job in fitting the wings of the PSF and produces a better wavelength solution overall. Note that a full wavelength calibration is computationally intensive. We recommend using the parallelization option for faster computing time (Ex. $\sim$ 10 min. using 4 cores).

\item At the beginning of each science sequence, a short 30 second \textit{H} band Argon Arc lamp exposure should be taken at the elevation of your target. This will be later used to correct for any shifts of the lenslets on the detector due to internal flexure of the IFS. \textit{H} band data are used to limit exposure time, and can also be used to provide an offset to other bands. 

\item These short arc lamp images should be reduced using the \textbf{Quick Wavelength Solution} Recipe in the GPI DRP. Instead of computing a solution for each lenslet individually, this recipe computes the locations of every 20th lenslet (by default) and computes the average shift across the entire field of view.

\item When reducing a science sequence, the wavelength calibration is added in the \textit{Load Wavelength Calibration} primitive. The files generated in the steps above should be selected automatically, but the user can manually input a file as well. To correct for the internal flexure of the IFS, we recommend using the `BandShift' mode in the \textit{Update Spot Shifts for Flexure} primitive. This primitive will use the short arc lamp image to correct for any shift in the lenslet spectra that occurred between the high SNR wavelength calibration images and the science data. 

\end{itemize}

A description of each the steps above is available in the GPI pipeline documentation (\url{http://docs.planetimager.org/pipeline/}). In the next section, we discuss how to check the quality of a wavelength calibration file, and several suggestions for troubleshooting. 

\subsection{Quality Checks and Troubleshooting}

For the general GPI user, we have included several tools to inspect the quality of the wavelength calibration in the GPI DRP. Before beginning to reduce any arc lamp image files, we recommend inspecting them by hand. There is an issue in the communication software between the GCAL unit and GPI that occasionally causes a shutter to remain closed, allowing no photons to reach the instrument during an arc lamp exposure. Once you have confirmed that the raw files have counts, generate a wavelength calibration file using the steps outlined above. To check if the produced wavelength calibration file (extension `wavecal') is a good fit to your science data, display both images using GPItv (display module used in the GPI DRP). First, open your science image in GPItv, and overplot the wavelength calibration file first by selecting it and then by using the `Plot Wavecal/Polcal Grid' in the Labels menu of GPItv. This will draw a grid of lenslet spectra over-top of the observed spectra. A window allows the user to manually adjust the x and y positions of the lenslet spectra before drawing. Alternatively, the `Move Wavecal Grid' Mouse Mode will allow the user to click and drag in the GPItv window to move the grid of lenslet spectra. 

The \textit{Quality Check Wavelength Calibration} Primitive is included in all Wavelength calibration recipes. It performs several checks that are often performed by eye in GPItv. First, the size of the wavelength calibration file is checked for the correct dimensions, and the primitive confirms that all of the lenslet positions are not the same. Histograms are generated for the x and y positions, dispersion ($\mu$m/pixel), and the tilt (orientation of the spectra on the detector in radians) of the lenslet spectra. If any two neighboring lenslets have values that are some threshold above the mean difference, the wavelength calibration file will fail the basic quality check. By default, this occurs if the x or y positions vary by more than 2 pixels, the dispersion varies by more than $0.002$ ~$\mu$m/pixel, or the spectral tilt varies by more than 0.1 radians ($\sim 5^{\circ}$). As the name implies, this is only a basic quality check. It is still recommended that the users display the wavelength calibration file overplotted with their science data in GPItv.

\section{Stability of the Wavelength Calibration}
\label{sec:stability}

After more than two years of on-sky science operations, we have a better understanding of how the stability of the wavelength calibration for the IFS changes with environment and time. Including both long exposure arc lamp images taken once per GPI run and short arc lamp images taken with each science sequence, we have generated over 1000 wavelength calibration files across all bands ($\sim$ 800 in \textit{H}-band and 50 - 100 in each of the other filters), with data obtained as early as November 2013. 

We examined the behavior of the lenslet positions, spectral dispersion, and tilt with several header keywords relating to the environment on Gemini South including the temperature in the IFS and the detector temperature, the telescope elevation, and the date of the observation. We found no correlation with temperature inside or outside of the instrument. While the tilt of the lenslet spectra varies across the field of view, for a given lenslet the tilt is very stable with variations $< 10^{-3}$ radians in the entire sample. 

\begin{figure}
\begin{center}
\includegraphics[width=0.8\textwidth]{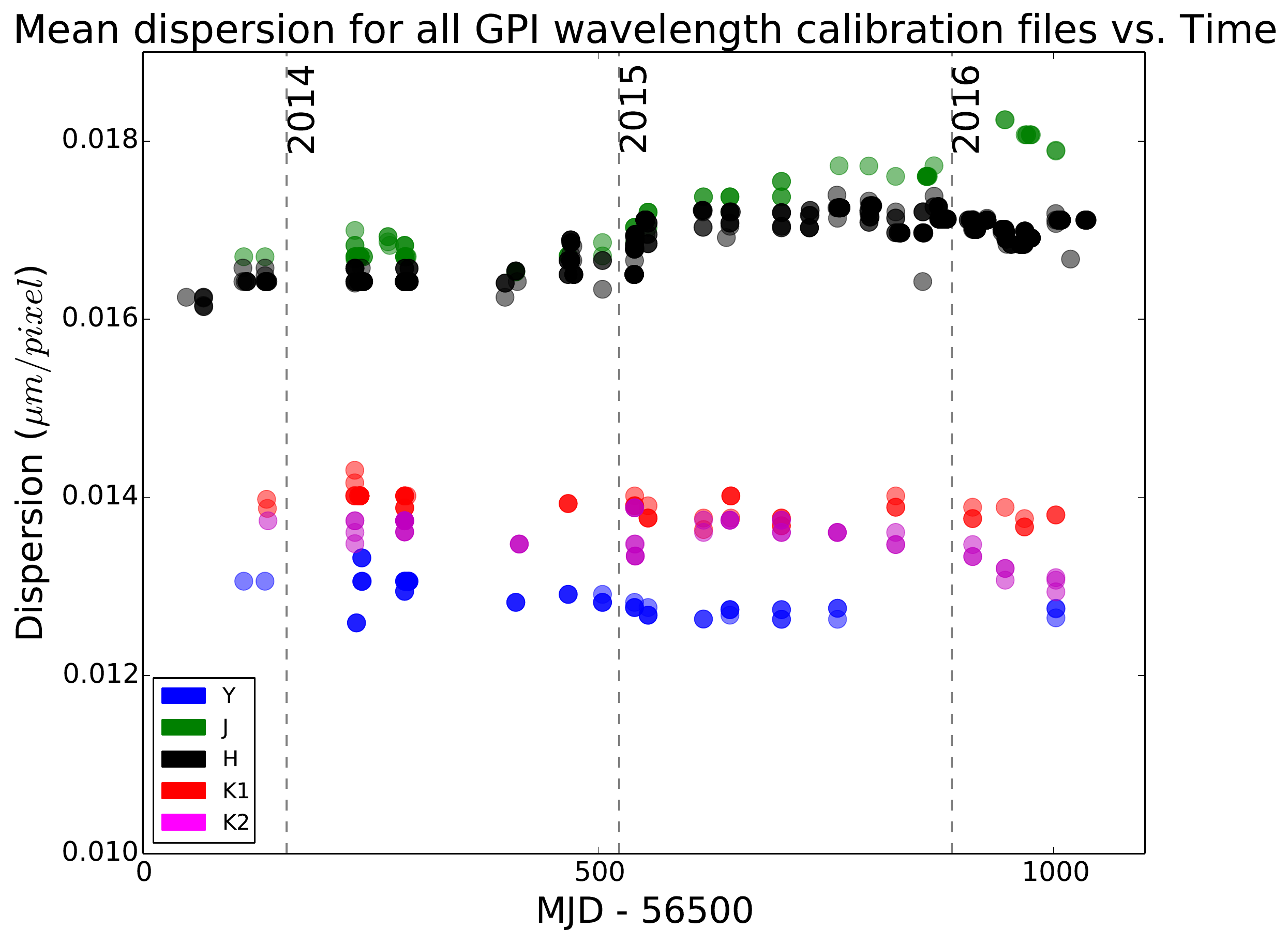}
\end{center}
\caption[example]
{\label{fig:dispersion} Mean spectral dispersion for all GPI wavelength calibration files as a function of time; colorcoded by filter. Beginning in 2015, there is a change in the spectral dispersion with time by $\sim 4 - 5$ \% over 1.5 years. }
\end{figure}

The spectral dispersion was observed to change with time gradually by $\sim 4 - 5$ \% in 1.5 years, with an apparent turning point at the beginning of 2015, and potentially another change at the beginning of 2016 (Figure \ref{fig:dispersion}). The direction and magnitude of the change varies with band, and the reason for this evolution remains unknown. It is unclear whether this is a gradual or discontinuous change. Some number of jumps with superimposed noise could mimic a gradual change in the data, and may be easier to explain with a mechanical change in the IFS. It is possible that a mechanical change in the prism could cause some angle to drift over time, but we have not identified an optic in the IFS that could reproduce this change in dispersion while conserving the tilt of the lenslet spectra. 

Alternatively, the dispersion change could be an artifact of the aging of the GCAL arc lamps. Any changes in excitation associated with aging cathodes, gas contamination or leaks etc. could effect our ability to distinguish between the blended lines that make up the low resolution GPI spectra. This could explain why the effect is largest in \textit{J} and \textit{H} bands which have the fewest distinct spectral lines. We see no change in the trend between Xe and Ar lamp data, though we only have Xe arc lamp data in \textit{H}-band. This could also be a data processing effect. We have not reduced all of the wavelength calibration files with the same version of the Data Reduction Pipeline. However, when re-processing a subset of older calibration files, we found no change in the dispersion.

As mentioned previously, the x and y positions of the lenslets vary with flexure. The internal flexure of the IFS has two components: (1) Repeatable sub-pixel shifts vary with the elevation of the telescope as the gravity vector on the IFS changes. This behavior is well defined by a hysteresis curve and can be easily corrected. (2) Occasionally, a large shift of $\leq \, \pm 3$ pixels in either x or y will result from the shifting of an unidentified optic within the IFS by a few millimeters. This is unpredictable and must be corrected using 30 second arc lamp images taken before each science sequence. 

\begin{figure}
\begin{center}
\includegraphics[width=0.8\textwidth]{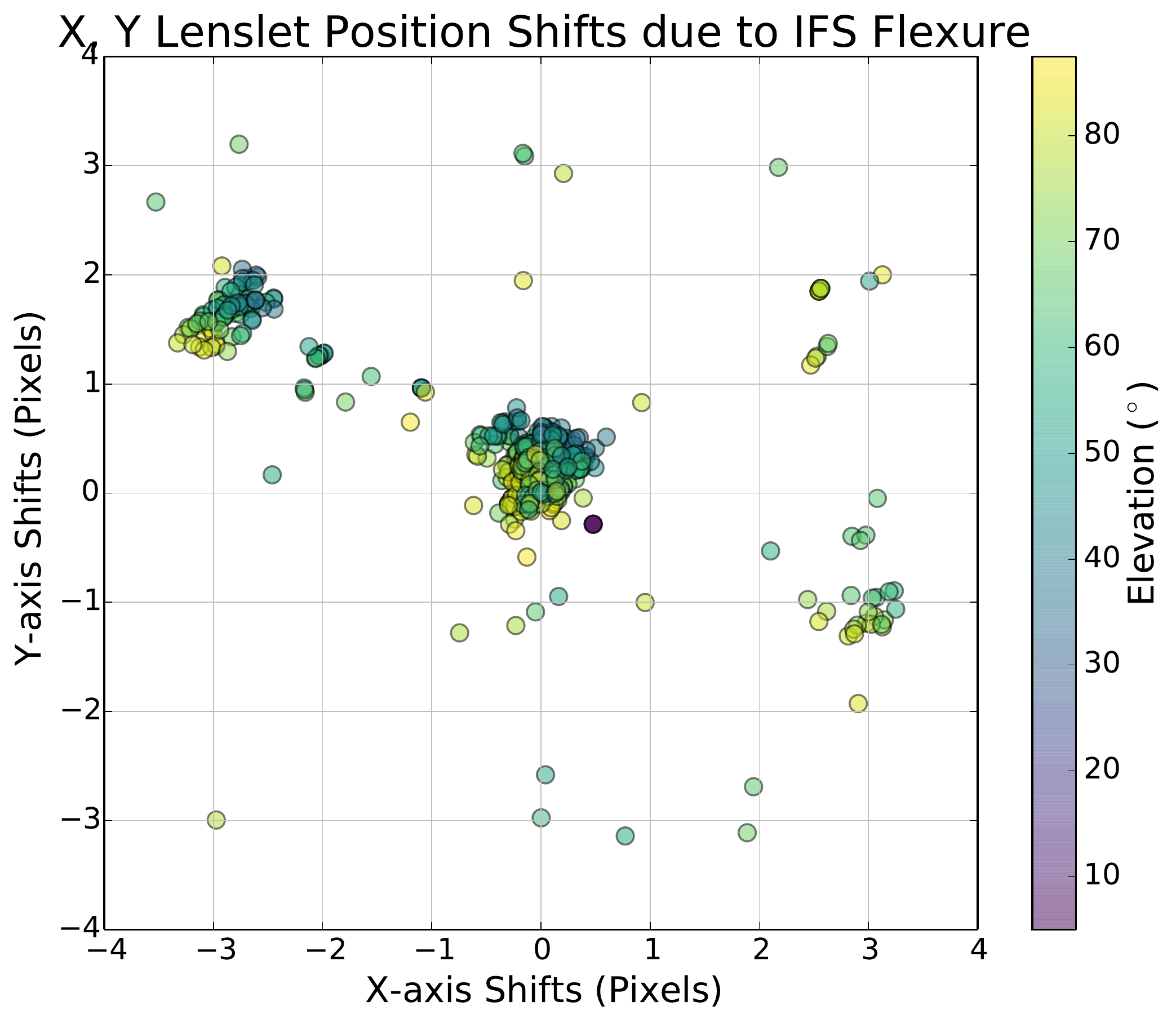}
\end{center}
\caption[example]
{\label{fig:flexure} The change in the x and y pixel locations for the central lenslet between the quick arc lamp image, and the long exposure arc lamp image taken at the beginning of the run in the same bandpass. Color coded with elevation. Zenith is at 90$^{\circ}$}
\end{figure}

Both behaviors are demonstrated in Figure \ref{fig:flexure}, which gives the change in the x and y pixel locations for the central lenslet between the short arc lamp image (quick), and the long exposure arc lamp image (master) taken at the beginning of the run in the same bandpass, color coded with elevation. While there is some scatter, most points in the sample fall into three distinct regions. This is likely a result of the loose optic oscillating between two states, i.e. state (a) and state (b). For the largest population, both the master and quick wavelength calibration files were taken with the optic in the same state. The two other groups correspond to occasions when the master files were taken with the optic in state (a) and the quick files were taken with the optic in state (b) and vice versa. Within each of the three populations, there is a clear trend in elevation. Larger shifts occur as the telescope is moved away from zenith (90$^{\circ}$).

\section{Nonlinearity of the Wavelength Solution}
\label{sec:nonlinear}

In this section, we investigate any departures from linearity of the wavelength solution. We examine any uncertainties in the spectral positions that could result from the assumption of a linear dispersion solution, and look for alternate expressions for the wavelength as a function of position on the detector. We first use simulations of the IFS output on the detector, and compare this to GCAL arc lamp observations. We also use GPI's Non-Redundant Mask mode for an independent measurement of the wavelength scaling. 

\subsection{Modeled Performance}

Using the geometry of the optical path within the GPI IFS, we are able to examine the theoretical spectrum of each lenslet produced by the dispersing prism. We derived the theoretical pixel position as a function of wavelength, $\lambda$, by combining the optical path of the GPI IFS using Zeemax files with the Sellmeier approximations for the indices of refraction of both glass components of the dispersing prism. This was then propagated into pixel values using an estimate for the effective camera focal length of 232 mm and the pixel size of 18 $\mu$m. 

\begin{figure}
     \centering
      \subfigure[Linear Dispersion]
	{
     \includegraphics[width=0.4\textwidth]{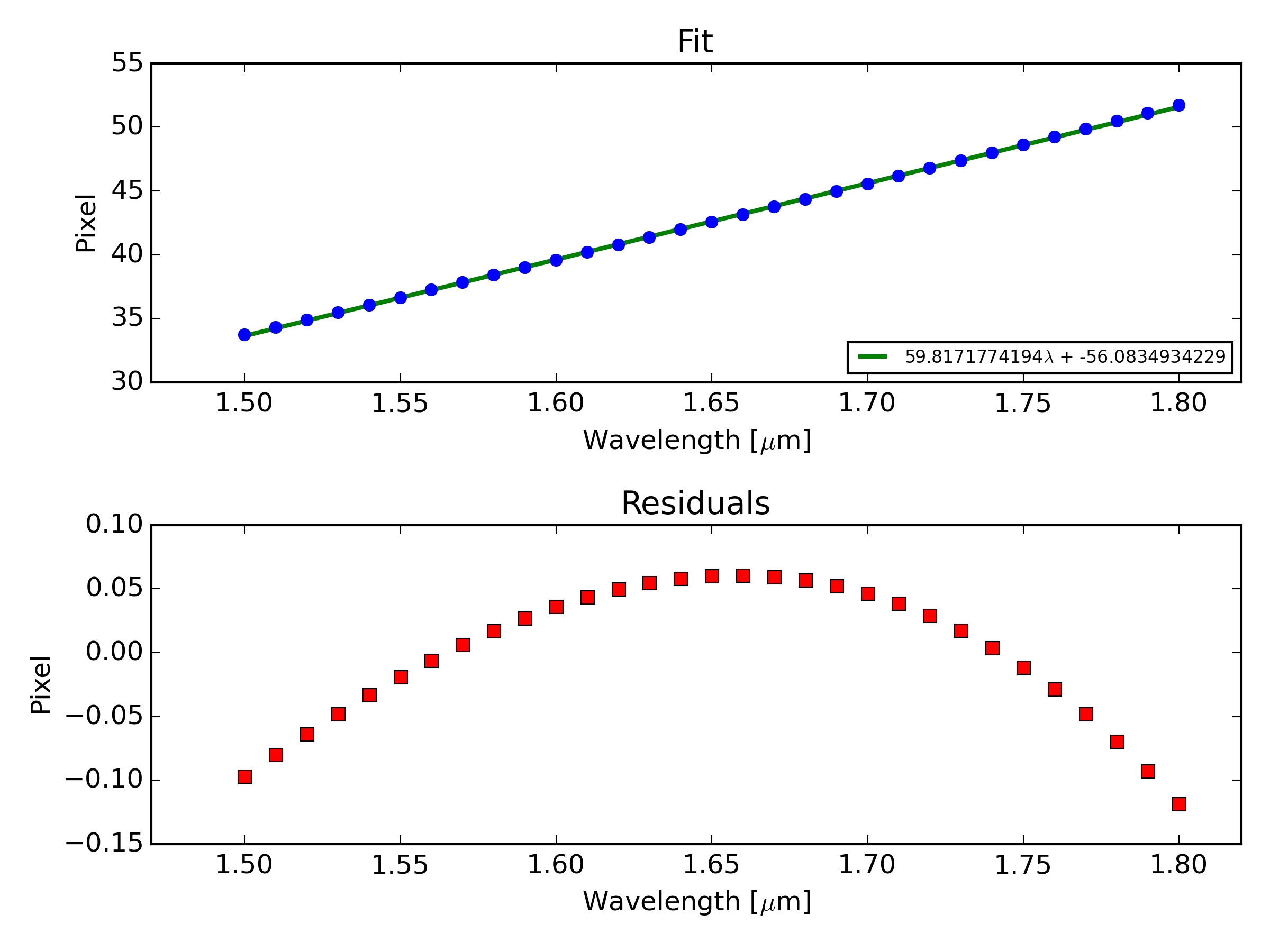}
     \label{fig:zlin}
  	 }
   \subfigure[Quadratic Dispersion]
   {
   \includegraphics[width=0.4\textwidth]{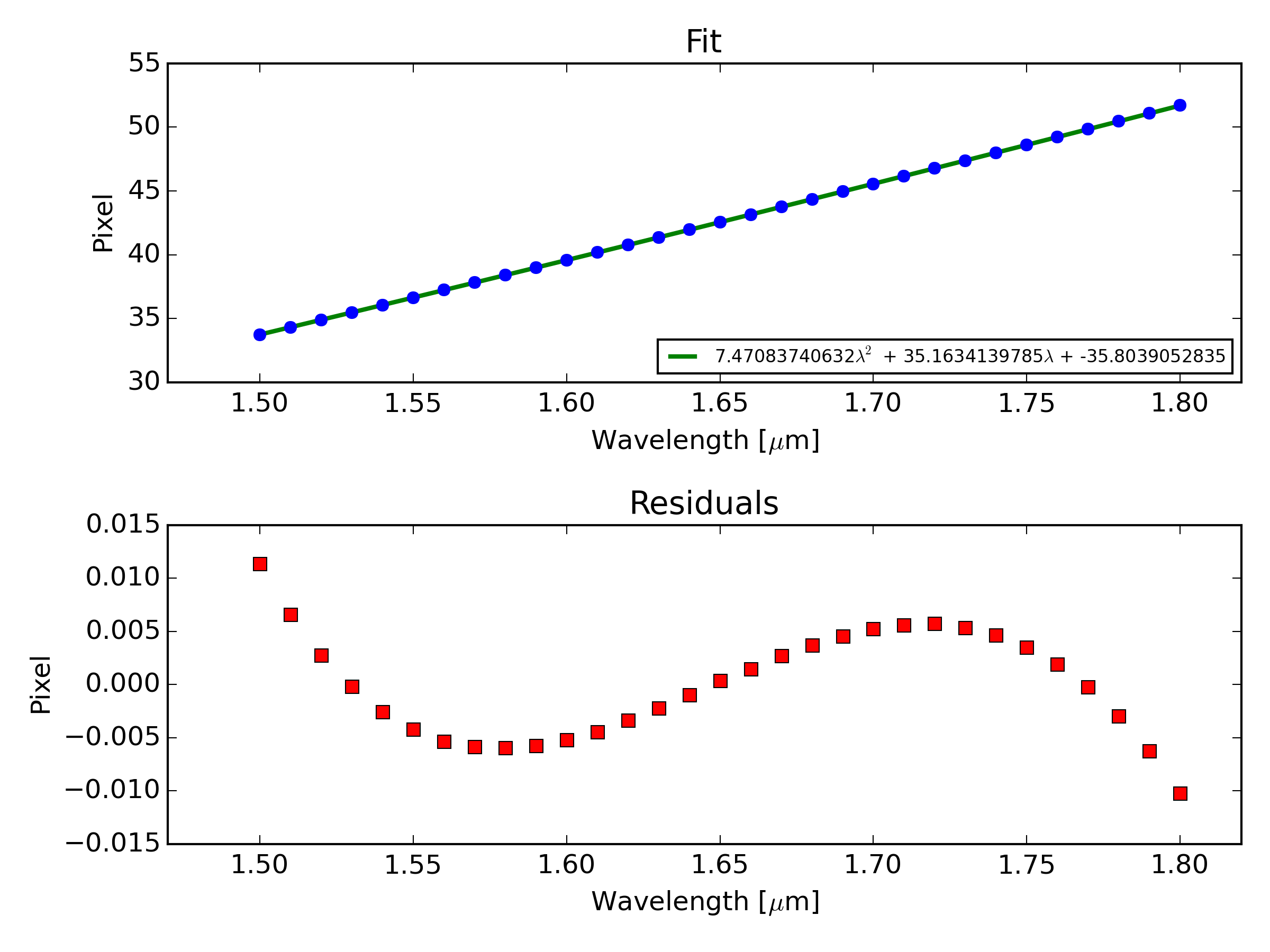}
   \label{fig:zquad}
   }

   \caption[example] 
   { \label{fig:zmaks} Theoretical fits to the dispersion of the GPI IFS usng (a) a linear model for the position as a function of wavelength, and (b) a quadratic model. The quadratic solution results in errors that are an order of magnitude less than the linear case with uncertainties in the pixel positon of $< 0.01$ pixels. }
   \end{figure}

We fit the resulting dispersion by several n-dimensional polynomials in $\lambda$ with n = 1 - 4 for each band individually. In the case of a 4th degree polynomial, the residuals no longer show any obvious structure. However, given our ability to resolve spectral features on the detector plane, it is unlikely that we will be able to detect pixel positons below the 1/100th of a pixel level. A quadratic solution for position as a function of $\lambda$ is required for uncertainties $< 0.01$ pixels. This gives an order of magnitude improvement in pixel uncertainties from a linear fit to the dispersion. An example for \textit{H}-band is given in Figure \ref{fig:zmaks}, and this behavior is demonstrated in all filters.

\subsection{Non-Redundant Masking as an Independent Wavelength Check}

GPI is equipped with a Non-Redundant Mask (NRM); an interferometric mask containing 10 holes separated by non-redundant baselines.\cite{2014SPIE.9147E..7BG} GPI's NRM mode complements GPI's coronagraph by accessing a smaller Inner Working Angle (R $\sim \lambda$/2D where D is the longest baseline) at moderate contrast. The size of the NRM PSF directly depends on the wavelength and is a sensitive probe of changes to pixel sampling with changing wavelength. Consequently, the GPI NRM allows for wavelength calibration of the GPI IFS without requiring an external calibration source.


An independent measure of the wavelength (assuming a pixel scale)  can be accomplished by cross-correlating NRM exposures with simulated PSFs at varying pixel scales. Reference PSFs can be generated numerically knowing the pupil geometry and varying the wavelength scaling around an expected value for the data. While the absolute wavelength calibration depends also on the pixel scale, the NRM can measure deviations from a linear wavelength function across a given filter. For each spectral channel, the NRM PSF is cross-correlated with each of the reference PSFs, and the data are fit to a parabola to determine peak correlation. The location of the peak corresponds to the best fit magnification for that spectral channel. Figure 4 shows the accumulation of several NRM datasets in various bands, comparing the measured wavelength compared with the wavelength reported by the GPI pipeline solution. The residuals are generally 0.05 μm which agrees well with what is predicted from the modeled performance and these errors could likely be reduced by an order of magnitude. Re-reducing the NRM images with updated non-linear corrections could help verify that a new solution is accurately representing the true dispersion.

  \begin{figure}
     \centering
      \subfigure[]
	{
     \includegraphics[width=0.45\textwidth]{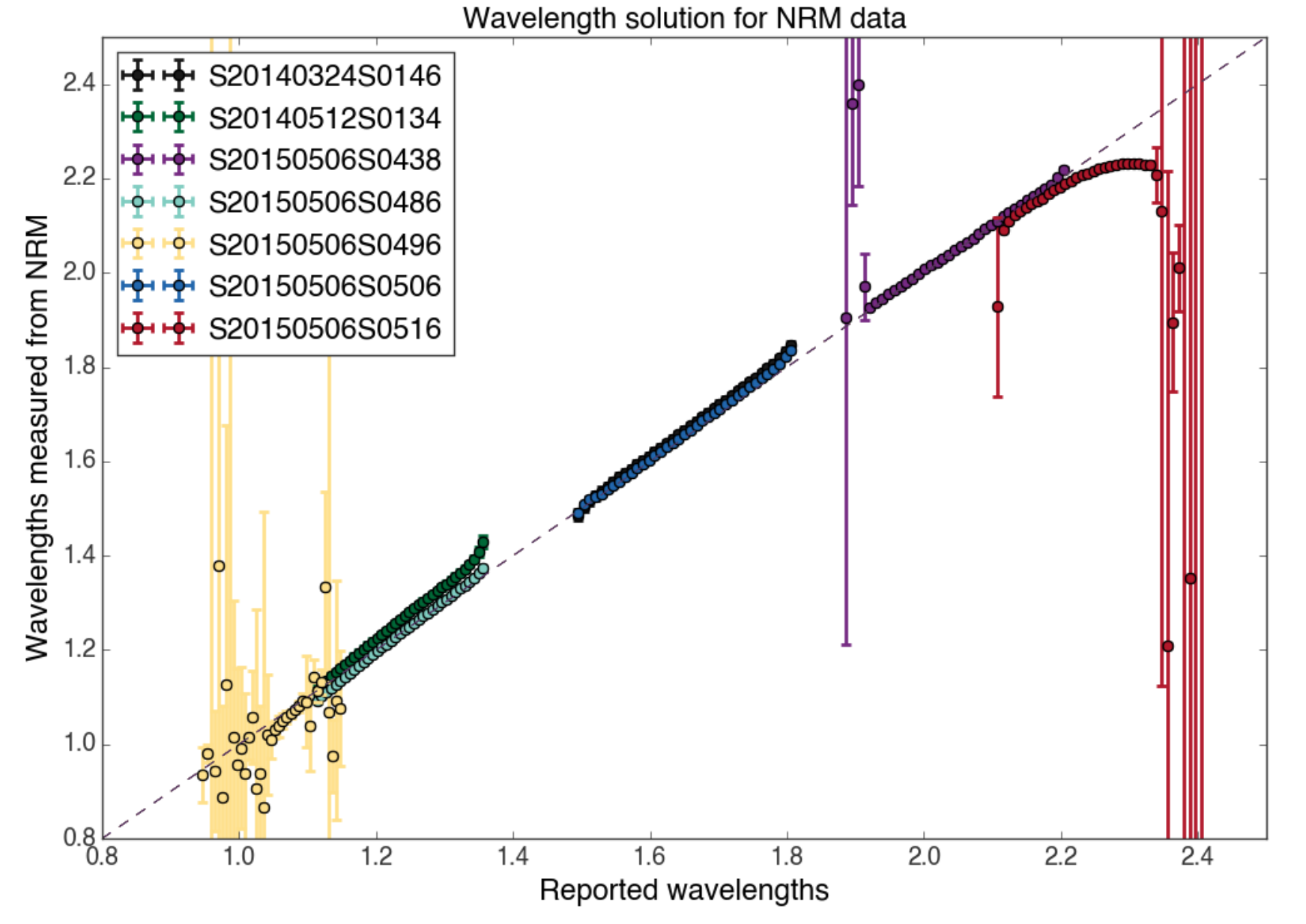}
  	 }
   \subfigure[]
   {
   \includegraphics[width=0.45\textwidth]{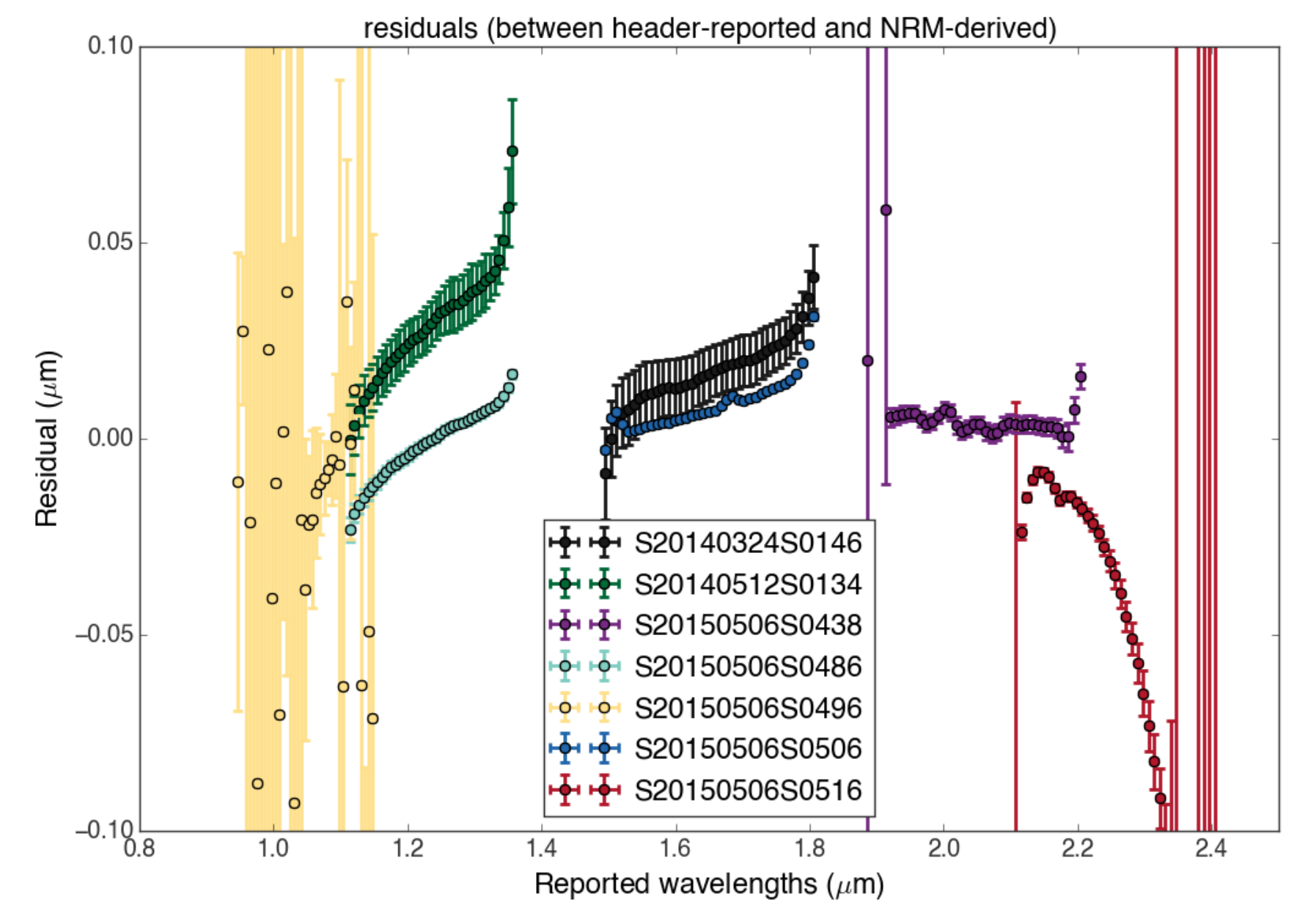}
   }
   \caption[example] 
   { \label{fig:nrm} (a) Wavelength solution derived using the NRM pupil for several datasets.  (b) Residuals from a linear fit to the wavelengths derived from the arc lamp wavelength calibration to the wavelengths found in the fit the NRM PSF. All plotted datasets were taken before cryocooler upgrades and suffer vibrations, which blur out fringes. This is especially degrading for Y band data. Lower throughput in K2 could have contributed to the large error-bars measured. }
   \end{figure}

\subsection{Quadratic Wavelength Solution}

With the 2D wavelength calibration algorithm, we provide a wavelength solution accurate to 1/10th of a pixel. This sensitivity allowed us to examine the departure of the IFS prism dispersion  from linearity. The x and y pixel position for each lenslet is defined by a unique $x_{0}$ and $y_{0}$ position for a reference wavelength $\lambda_{0}$ (which varies with filter), the tilt ($\theta$, angular orientation of the lenslet in radians from vertical), and a dispersion. In the linear case, the dispersion is assumed to be linear (with coefficient $w$), and the positions are defined by: $x = x_{o} + \sin{\theta} \frac{\lambda - \lambda_{o}}{w}$ and $y = y_{o} - \cos{\theta} \frac{\lambda - \lambda_{o}}{w}$ where the wavelength, $\lambda$, for a given pixel is given by $\lambda = w \sqrt{ ( x - x_{0} )^{2} + ( y - y_{0} )^{2}} + \lambda_{0}$. For the quadratic formalism of dispersion, we expand these equations in ($\lambda - \lambda_{o}$) as shown in the Equation below.

\begin{equation*}
x = x_{o} + \sin{\theta} \bigg[ ( \lambda - \lambda_{0}) / w + \boldsymbol{q ( \lambda - \lambda_{0})^{2}  }\bigg] \,\,\,\,\,\, \mathrm{and} \,\,\,\,\,\, y = y_{o} - \cos{\theta} \bigg[ ( \lambda - \lambda_{0}) / w + \boldsymbol{q ( \lambda - \lambda_{0})^{2} } \bigg] 
\label{eq:lambda}
\end{equation*}

The quadratic term is shown in bold. After some experimentation, the coefficient, $q$, was found to be well fit by a value of $w/10$. To limit computation time, we fix this value when determining the wavelength solution for each lenslet. The magnitude of the quadratic component varies with filter, but is generally less than 0.003 \% of the linear term (ie. $\frac{q}{1/w} <  3 \times 10^{-5}$), and this serves as a small perturbation to the linear case. 

The largest improvement was seen in the reduction of the \textit{H}-band. A comparison between the linear and quadratic fits for a single lenslet in a Xe arc lamp dataset is given in Figure \ref{fig:res}. The quadratic solution improved the reduced $\chi^{2}$ by a factor of $\sim$ 2 above the linear case (both using the microlens PSF), and gives an position accuracy of 0.012 pixels, which is approaching the theoretical limit of 0.01 pixels (Figure \ref{fig:hist}). In other bands, the improvement is less pronounced. In \textit{K1} band for example, the quadratic solution only provided a factor of 1.05 improvement over the linear case. This is likely a consequence of our inability to resolve spectral lines in the Argon arc lamp data, rather than a failure of the quadratic dispersion solution to accurately define the data. While the Argon arc lamp available at Gemini South provides the best SNR with the shortest exposure times, the spectral lines in all filters are blended and difficult to distinguish at the resolution of GPI. The Xenon lamp provides better distinction between spectral lines in some filters, but the count rate of the lamp is prohibitively low (Ex. Ar exposure for K2 bands takes $\sim$ 2 hours, and Xe would require at least twice the exposure time to achieve the same SNR). 

   \begin{figure}
     \centering
      \subfigure[]
	{
     \includegraphics[width=0.13\textwidth]{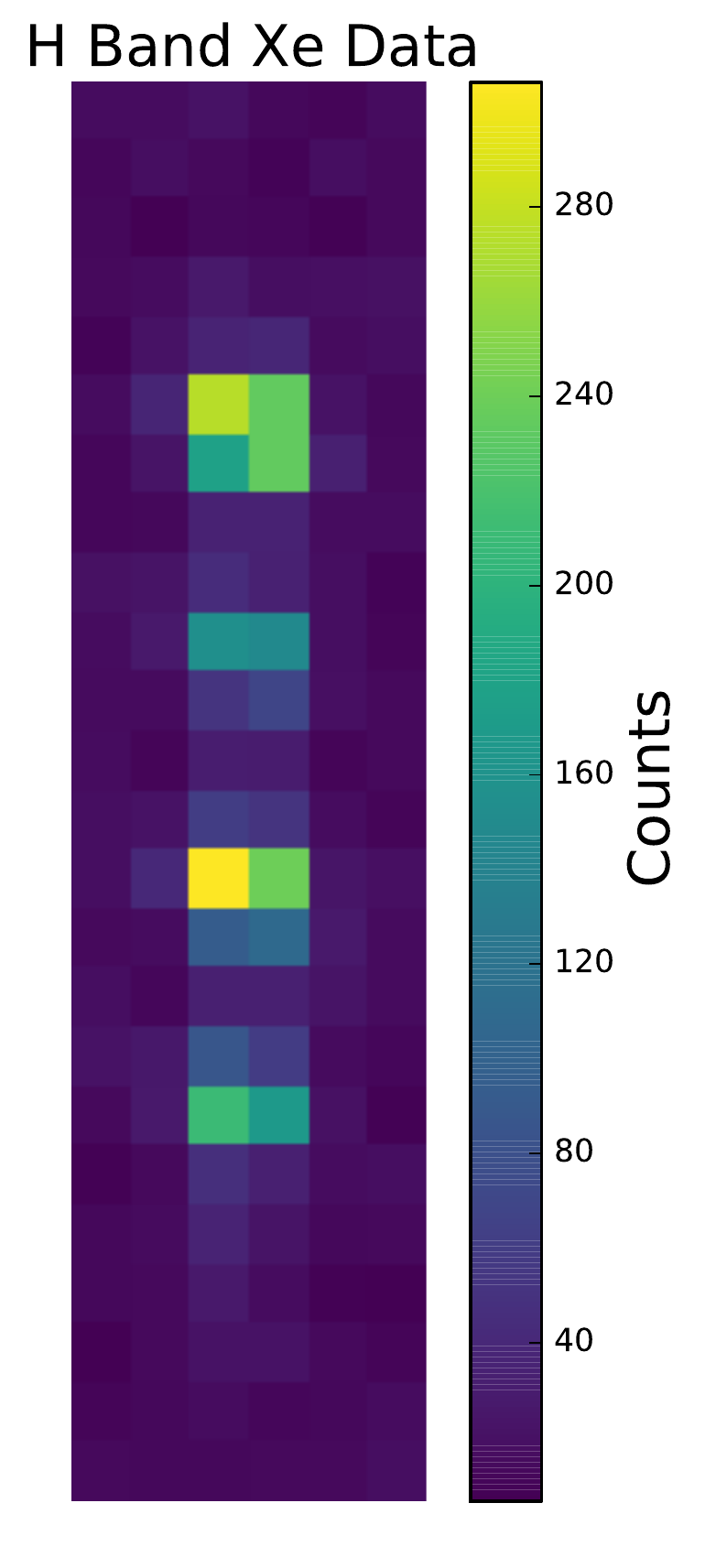}
     \label{fig:gaussres}
   }
   \subfigure[]
   {
   \includegraphics[width=0.13\textwidth]{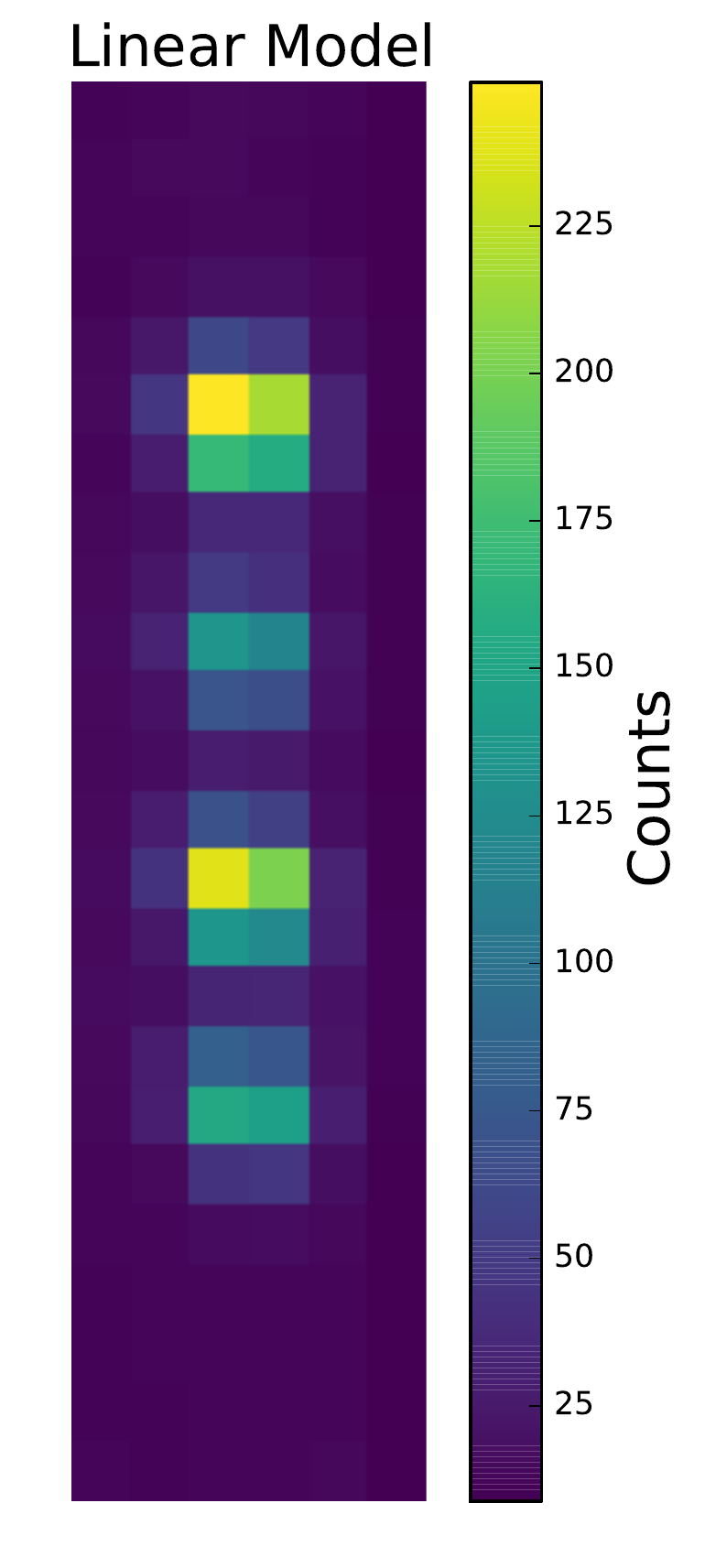}
   \label{fig:microlensres}
   }
   \subfigure[]
    {
   \includegraphics[width=0.13\textwidth]{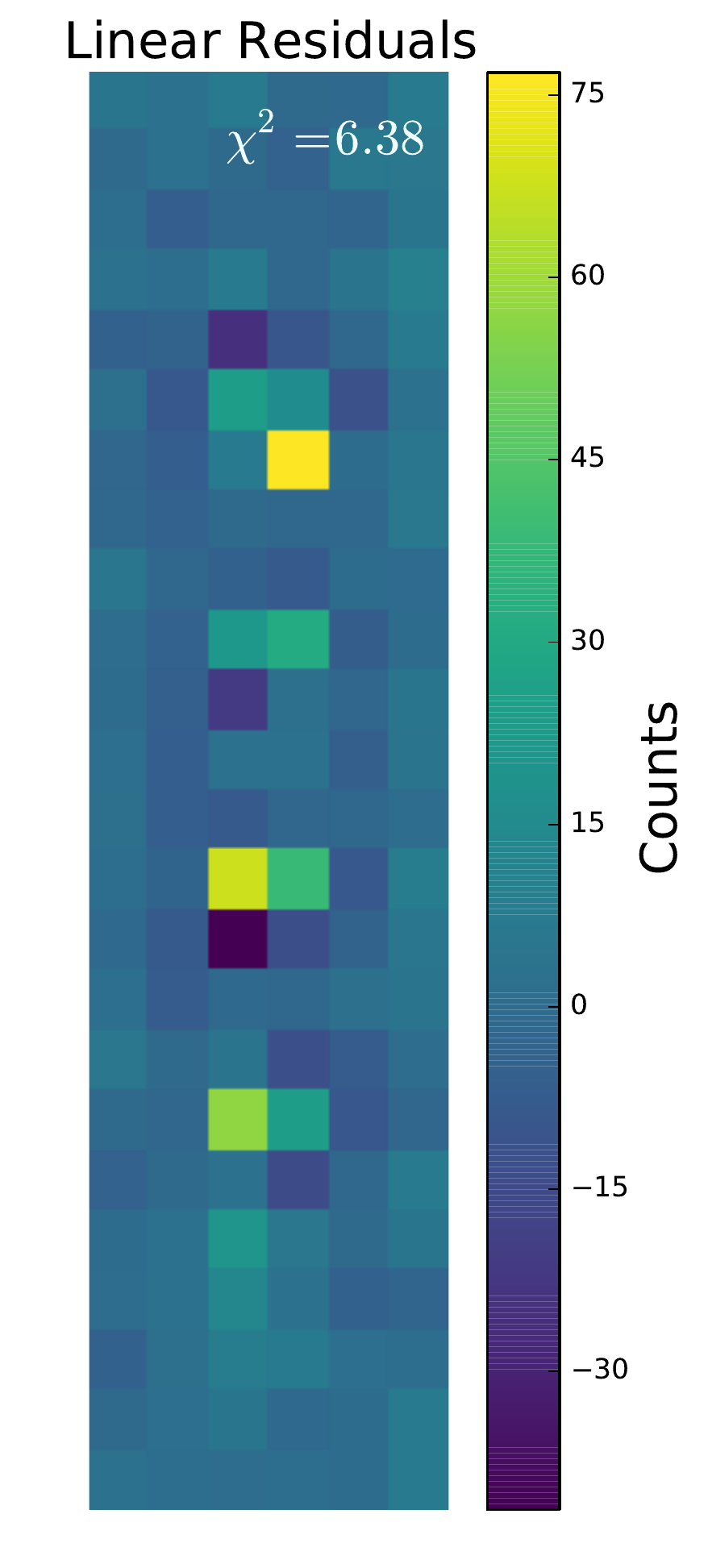}
   \label{fig:microlensres}
   }
      \subfigure[]
   {
   \includegraphics[width=0.13\textwidth]{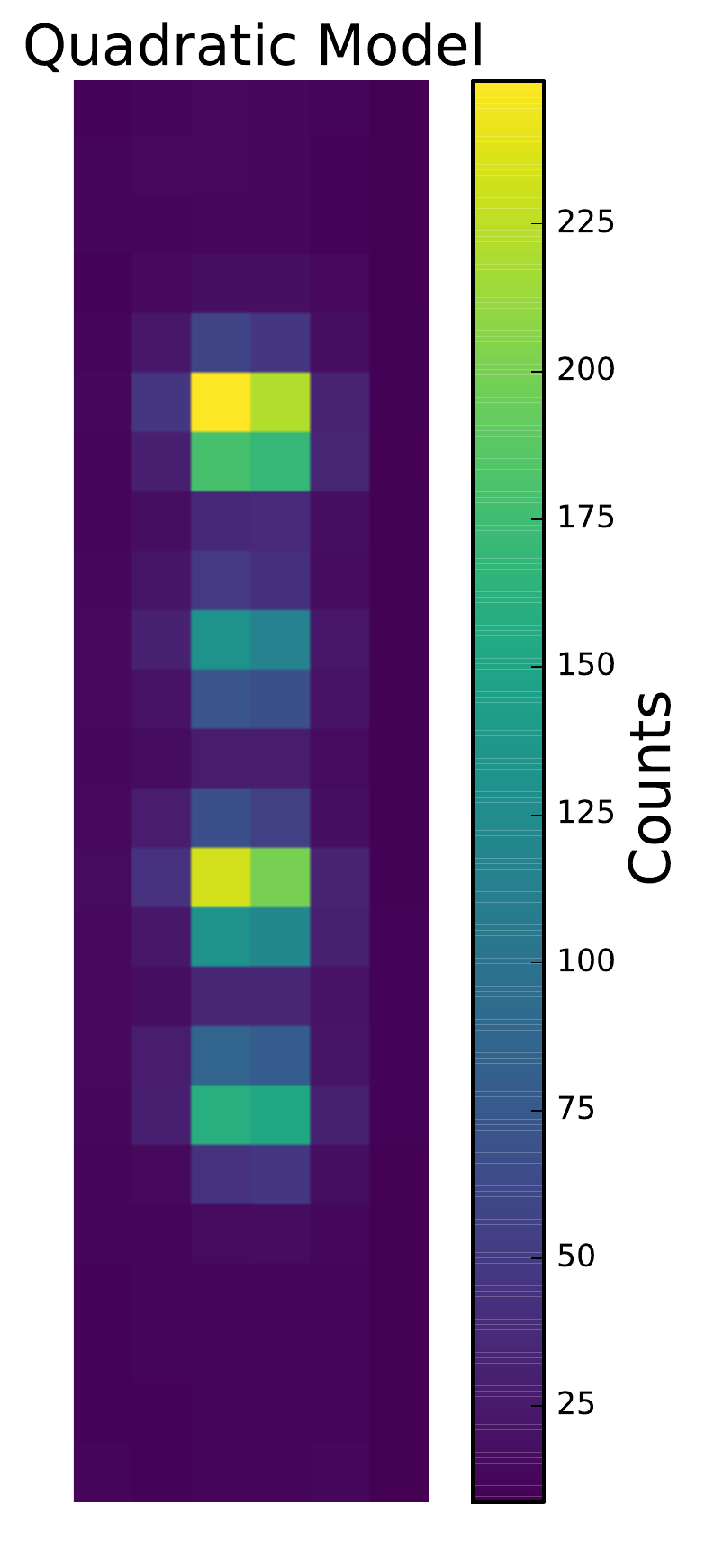}
   \label{fig:microlensres}
   }
   \subfigure[]
    {
   \includegraphics[width=0.13\textwidth]{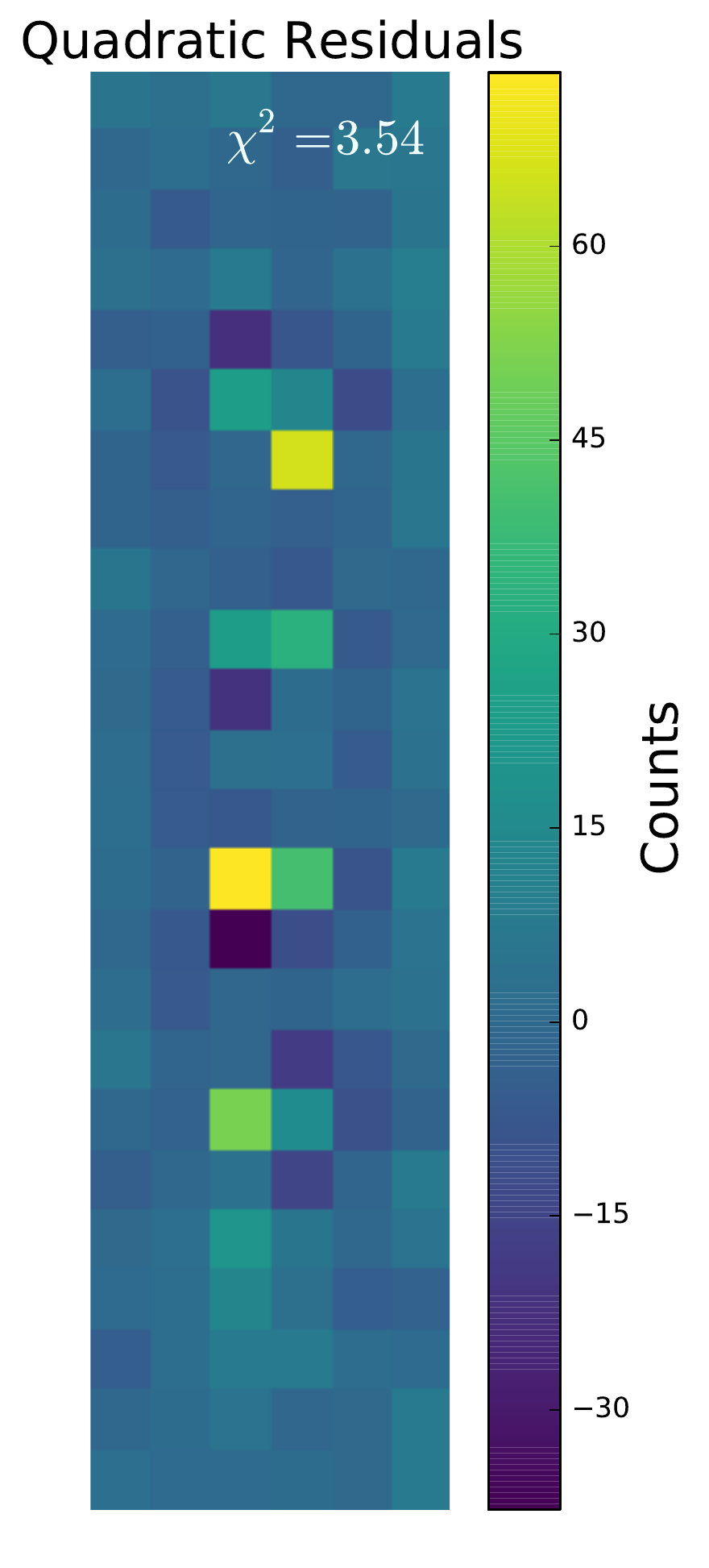}
   \label{fig:res}
   }
      \subfigure[]
   {
   \includegraphics[width=0.13\textwidth]{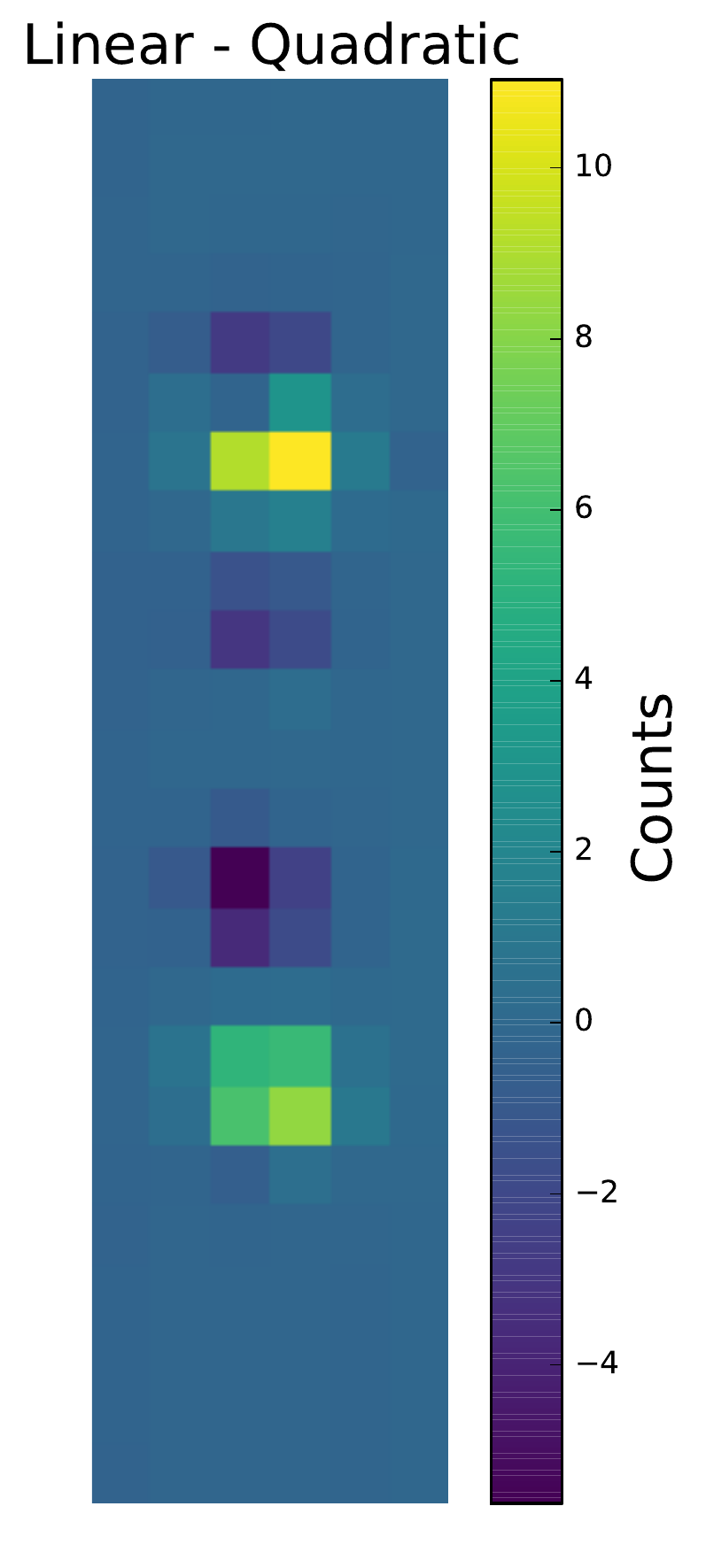}
   \label{fig:res}
   }
   \caption[example] 
   { \label{fig:residuals} 
 (a) A single lenslet cutout of an observed H band Xe arc lamp image. (b) A model Xe arc lamp image created using a linear dispersion model. (c) The residuals obtained by subtracting the observed lenslet spectrum array from the linear dispersion modeled lenslet array. (d) A model Xe arc lamp image created using a quadratic dispersion model.  (e) The same as (c) for the quadratic dispersion model, and (f) shows the linear model subtracted by the quadratic model. Note the scale of the residuals. These are largely a consequence of  the microlens PSFs being too narrow in the peak than the actual instrumental PSF, and less a result of our uncertainty in the locations of the spectral peaks. }
   \end{figure}

  \begin{figure}
  \begin{center}
   \includegraphics[width=0.4\textwidth]{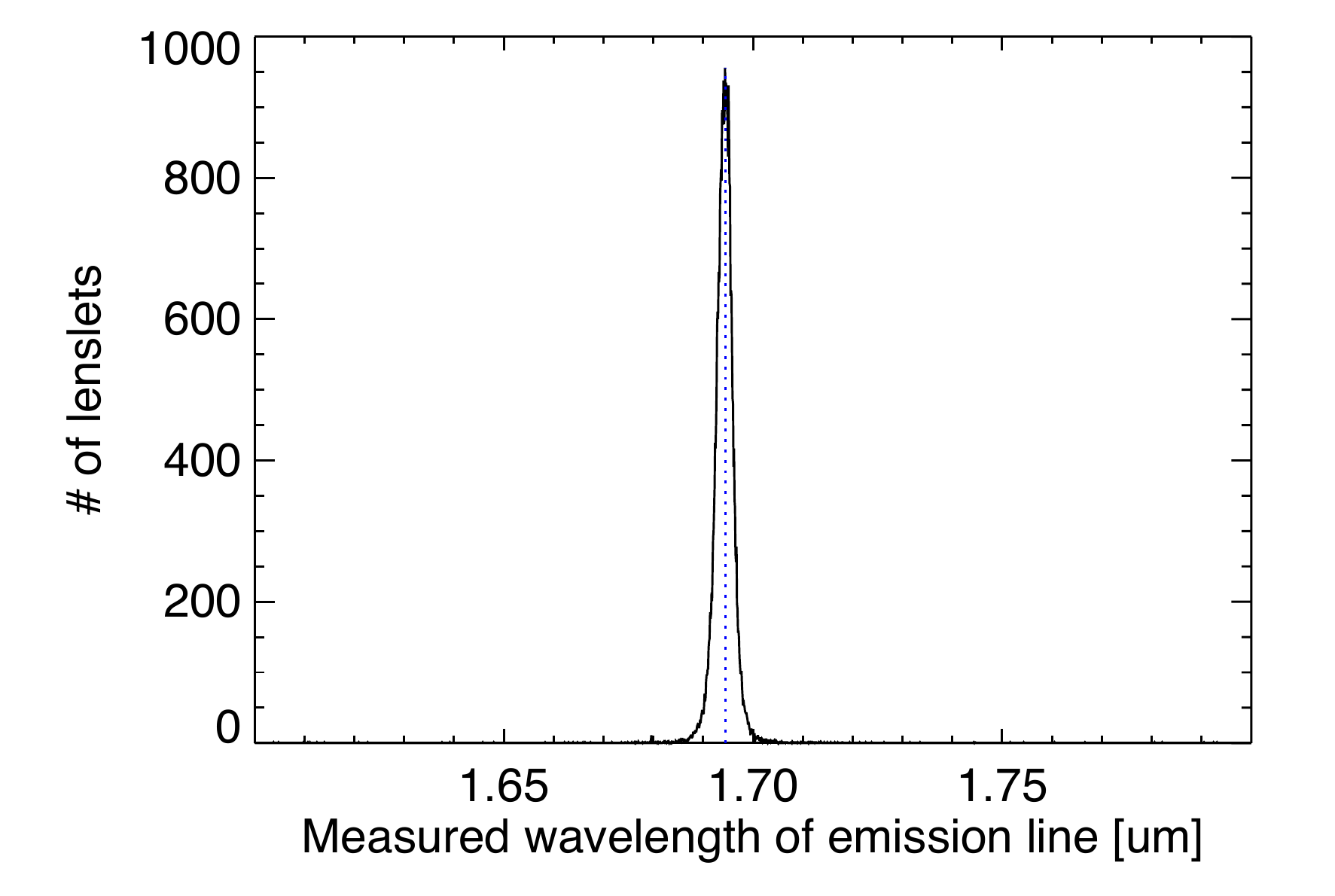}
   \end{center}
      \caption[example] 
   { \label{fig:hist} Histograms of the measured wavelength in $\mu$m of the spectral peak for an H band Ar image for all the lenslets in the image for a quadratic dispersion solution. The dotted line represents the theoretical location of the peak. The quadratic solution peaks at a value within 0.012 pixels of the correct wavelength.}
   \end{figure} 

\subsection{Simultaneous Wavelength Calibration of all Bands}

We have shown the improvement afforded by including a quadratic term to the wavelength solution for a single GPI filter, particularly in $H$-band. Furthermore, it should also be possible to fit the wavelength solutions over the full GPI bandpass simultaneously ($Y-K2$, corresponding to 0.9--2.4$\mu$m), assuming separate linear and/or higher-order fits to each of the individual bands. By defining wavelength vs. position across the full bandpass, this would provide small, yet non-negligible improvements to datacube extraction at different wavelengths. As described previously, quick $H$-band Argon arcs are taken as part of standard GPI observing sequences to calibrate observations at other bands, given the expediency of $H$-band arcs as opposed to the longer integrations required for arcs at different wavelengths, in particular $K1$ and $K2$. This bootstrapping process, with either a look-up table of flexure shifts, or in `BandShift' mode, involves extrapolating the shifts due to flexure from the $H$-band arcs to the band of the science observations, using the x- and y-positions from the most recent wavelength calibration. For this procedure, a higher fidelity spectral extraction may be possible if the extrapolation also incorporates a functional relationship between the wavelength solutions at different bands. 

\begin{figure}
\begin{center}
\includegraphics[width=0.8\textwidth]{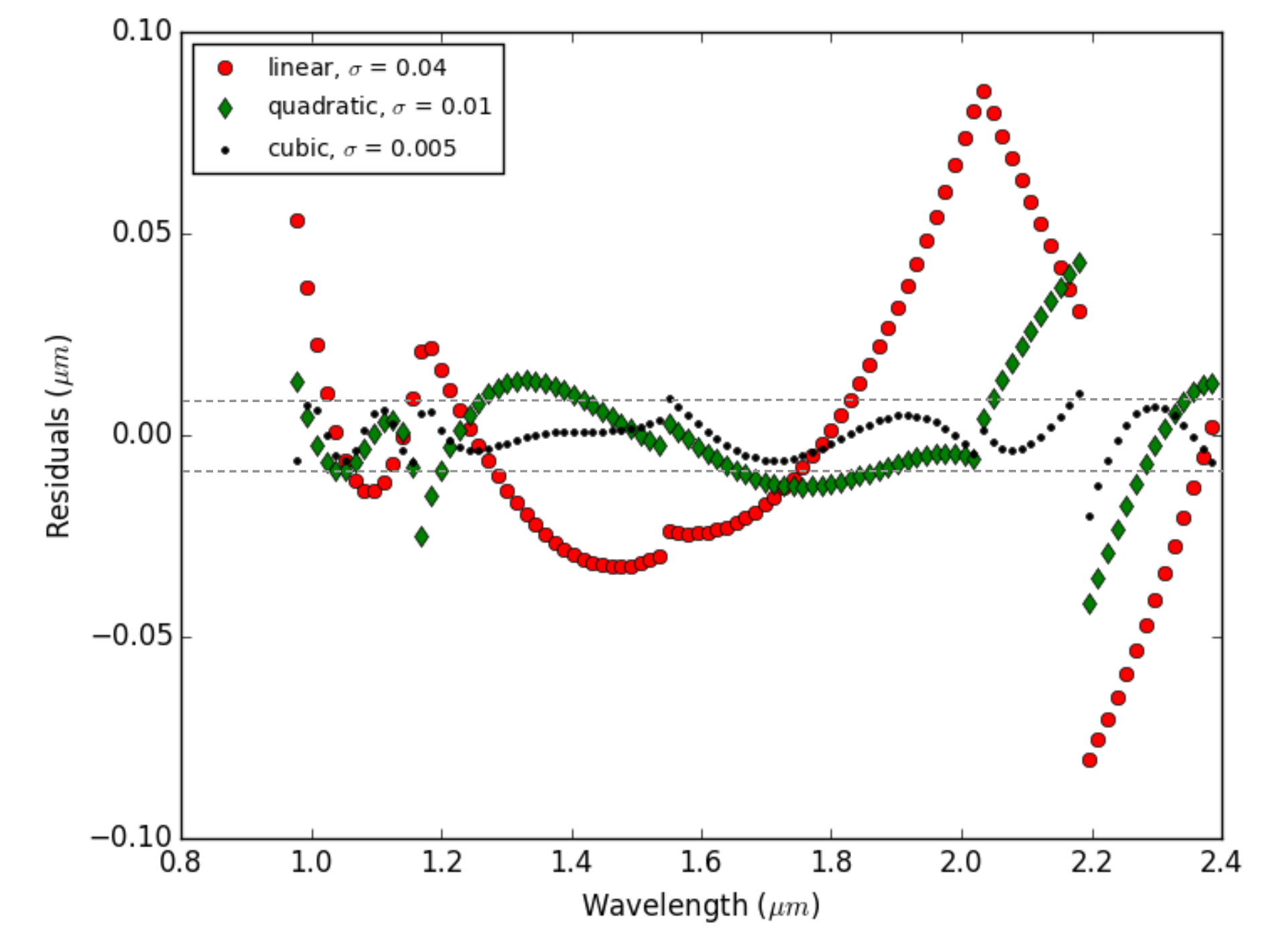}
\end{center}
\caption[example]
{\label{fig:fullfit} Residuals in $\mu$m from an overall polynomial fit to each of the independent linear solutions for the five GPI bands ($Y-K2$). The cubic polynomial fit over the full bandpass, shown with black circles, provides the minimal residuals to the wavelength solution in two dimensions, and will be implemented in efforts to improve the fidelity of using quick $H$-band arcs to calibrate data taken in other bands.}
\end{figure}

In an initial attempt to fit the full bandpass simultaneously, independent linear solutions for each of the five filters were stitched together and fit with a single polynomial, with results from using polynomials of different orders shown in Figure~\ref{fig:fullfit}. The residuals in $\mu$m from the different polynomial fits show that a higher-order polynomial, at least a cubic fit, is required to describe the full bandpass simultaneously in a self-consistent manner. Implementation of a multi-filter and full bandpass wavelength solution option into the existing GPI DRP wavecal primitives is ongoing, with future work involving comparison of the extracted datacube quality with science observations at different wavelengths.

\acknowledgments     
 
This material is based upon work supported by the National Science Foundation Graduate Research Fellowship under Grant Nos. DGE-1232825 (SGW, AZG) and DGE-1311230 (KWD) and by NASA/NNX15AD95G. Any opinion, findings, and conclusions or recommendations expressed in this material are those of the authors and do not necessarily reflect the views of the National Science Foundation. The Gemini Observatory is operated by the Association of Universities for Research in 
Astronomy, Inc., under a cooperative agreement with the NSF on behalf of the Gemini 
partnership: the National Science Foundation (United States), the National Research 
Council (Canada), CONICYT (Chile), the Australian Research Council (Australia), 
Minist\'erio da Ci\'encia, Tecnologia e Inova\c{c}\=ao (Brazil), and Ministerio de Ciencia, 
Tecnolog\'ia e Innovaci\'on Productiva (Argentina). 

\bibliography{report}   
\bibliographystyle{spiebib}   

\end{document}